\documentclass[twocolumn,floatfix,showpacs,amsmath,amssymb,pra]{revtex4}
\usepackage{times}
\usepackage{graphicx}
\usepackage{dcolumn}
\usepackage{bm}

\newcommand{\xv}{\vec{x}}
\newcommand{\yv}{\vec{y}}
\newcommand{\Tr}{\mathrm{Tr}}

\begin{document}

\title{Non-equilibrium dynamics of a Bose-Einstein condensate in an optical lattice\\ 
       in the 2PI $\mathbf{1/\cal N}$ approach}
\author{Kristan Temme}
\thanks{email:K.Temme@thphys.uni-heidelberg.de}
\author{Thomas Gasenzer}
\thanks{email:T.Gasenzer@thphys.uni-heidelberg.de}
\affiliation{Institut f\"ur Theoretische Physik, Universit\"at Heidelberg, Philosophenweg 16, 69120 Heidelberg, Germany}
\date{\today}

\begin{abstract} 
\noindent
The dynamical evolution of a Bose-Einstein condensate trapped in a one-dimensional lattice potential is investigated theoretically in the framework of the Bose-Hubbard model.
The emphasis is set on the far-from-equilibrium evolution in a case where the gas is strongly interacting.
This is realized by an appropriate choice of the parameters in the Hamiltonian, and by starting with an initial state, where one lattice well contains a Bose-Einstein condensate while all other wells are empty.
Oscillations of the condensate as well as non-condensate fractions of the gas between the different sites of the lattice are found to be damped as a consequence of the collisional interactions between the atoms.
Functional integral techniques involving self-consistently determined mean fields as well as two-point correlation functions are used to derive the two-particle-irreducible (2PI) effective action.
The action is expanded in inverse powers of the number of field components $\cal N$, and the dynamic equations are derived from it to next-to-leading order in this expansion.
This approach reaches considerably beyond the Hartree-Fock-Bogoliubov mean-field theory, and its results are compared to the exact quantum dynamics obtained by A.M.~Rey {\it et al.}, Phys.~Rev.~A 69, 033610 (2004) for small atom numbers.
\end{abstract}
\pacs{03.70.+k, 03.75.Kk, 03.75.Lm, 05.70.Ln, 11.15.Pg\hfill HD--THEP--06--11}
\maketitle
\section{Introduction}
\label{sec:intro}

The dynamical evolution of ultracold atomic quantum gases driven far out of equilibrium are a subject of intensely growing interest.
Precision measurement techniques for many-body observables have been and are being developed with vigourous effort.
This technology has triggered a strong demand for progress in theoretically describing non-equilibrium quantum many-body dynamics of strongly interacting systems beyond mean-field approximations.
Recent highlights of this development include, e.g., the variation and enhancement of the atomic interactions on the basis of Feshbach scattering resonances \cite{Stwalley1976b,Tiesinga1992a,Inouye1998a,Mies2000a,Donley2002a} as well as the achievement of strongly correlated regimes within optical lattices \cite{Jaksch1998a,Greiner2002a}.
With these techniques, e.g., a Bose-Einstein condensate can suddenly be brought out of equilibrium, so that mean-field approximations such as chosen in Gross-Pitaevskii and Hartree-Fock-Bogoliubov theory no longer suffice to describe its time-evolution.  

Mean-field approximations are valid as long as both, classical statistical and quantum fluctuations are small.
This is, in general, only the case in situations close to equilibrium.
But even close to equilibrium, fluctuations can play an important role on long time scales.
For example, the long-time evolution towards the Bose-Einstein equilibrium state relies on the fluctuations appearing beyond mean-field order in a quantum many-body description.

Moreover, kinetic descriptions beyond mean-field order which have been studied intensely in the past (in the context of cold atomic gases cf., e.g., Refs.~\cite{Proukakis1996a,Shi1998a,Gardiner1997a,Giorgini1997a,Proukakis1998a,Walser1999a,Gardiner2000a,Walser2000a,Rey2005a,Baier2005a} and references therein), usually neglect the initial dynamics directly after the change in the boundary conditions which drive the system out of equilibrium.
This shortcoming is cured in dynamical approaches in which coupled equations of motion for the correlation functions are derived to describe the time evolution starting from a specific initial state.
The built-up of correlations beyond mean-field order, in these equations, is usually taken into account by means of non-Markovian integrations over the evolution history of the correlation functions, see, e.g.~\cite{Calzetta1988a,Berges2002a,Bhongale2002a,Kohler2002a,Rey2004a,Gasenzer2005a}.

In this article we consider the time evolution of a Bose-Einstein condensate trapped in a one-dimensional lattice potential in the framework of the Bose-Hubbard model.
The emphasis is set on the far-from-equilibrium evolution in a case where the gas is strongly interacting.
This is realized by an appropriate choice of the parameters in the Hamiltonian.
In the non-equilibrium initial state one lattice well contains a Bose-Einstein condensate while all other wells are empty.
At NIST, such a system has recently been realized in experiment \cite{Peil2003a}.

We find oscillations of the condensed as well as the non-condensed fractions of the gas between the different sites of the lattice.
These oscillations are damped as a consequence of the collisional interactions between the atoms.
Our theoretical approach is based on functional integral techniques involving self-consistently determined mean fields as well as two-point correlation functions.
Specifically, the dynamic equations are derived from the two-particle-irreducible (2PI) effective action \cite{Luttinger1960a,Baym1962a,Cornwall1974a}.
Any approximations necessary in practice are made at the level of this action before the dynamic equations are derived by functional differentiation.
This ensures crucial symmetries like total particle number and energy to be automatically conserved.
The 2PI effective action has been used to compare various perturbative approximations with the exact result for the dynamics of a strongly correlated one-dimensional lattice Bose gas \cite{Rey2004a}.
The approach yields dynamic equations which constitute a proper initial value problem and therefore are in principle valid over all times, including initial and long-term evolution.
In the Markovian limit, in which their validity is restriced to certain intermediate times, the well-known kinetic theory descriptions including the Gross-Pitaevskii and Bogoliubov-de Gennes equations as well as Landau and Beliaev quasiparticle damping can be rederived from these dynamic equations \cite{Rey2005a,Baier2005a}.

In this article we consider a non-perturbative approximation which reaches substantially beyond the Hartree-Fock-Bogoliubov mean field theory.
This non-perturbative approach is based on a systematic expansion of the 2PI effective action in powers of the inverse number of field components ${\cal N}$ \cite{Berges2002a,Aarts2002b}. 
We point out that this $1/\cal N$ expansion of the 2PI effective action goes substantially beyond and is not to be confused with the standard $1/\cal N$ expansion of the 1PI effective action.
For example, it can be used to calculate critical exponents characterizing correlation functions in the vicinity of a phase transition, even in the limit of vanishing ${\cal N}$ \cite{Alford2004a}. 
The 2PI $1/\cal N$ expansion has recently been used to describe the dynamics of an ultracold atomic Bose gas far from equilibrium  \cite{Gasenzer2005a}.
We compare our results to the exact quantum dynamics also obtained in Ref. \cite{Rey2004a}  for small atom numbers.

The 2PI $1/{\cal N}$ expansion to next-to-leading order yields dynamic equations which contain direct scattering, memory and ``off-shell'' effects. 
It allows to describe far-from-equilibrium dynamics as well as the late-time approach to quantum thermal equilibrium.
Recently, these methods have allowed important progress in describing the dynamics of strongly interacting relativistic systems far from thermal equilibrium for bosonic \cite{Berges2002a,Berges2003b,Mihaila2001a,Cooper2003a,Arrizabalaga2004a,Aarts2006a} as well as fermionic degrees of freedom \cite{Berges2003a,Berges2004b}. 

Our article is organized as follows:
In Section \ref{sec:2PIEA} we recall the functional description of the quantum many-body dynamics and present the set of coupled dynamic equations for the mean field and the two-point correlation functions.
We discuss consequences of this distinction for the self energies which quantify the beyond-HFB contributions in the dynamic equations for the correlation functions.
In Section \ref{sec:NumRes} we present numerical results for the non-equilibrium dynamics in a one-dimensional lattice potential and compare these to the results from corresponding exact calculations.
Our conclusions are drawn in Section \ref{sec:Concl}.

\section{The 2PI effective action approach to the dynamics of a lattice Bose gas}
\label{sec:2PIEA}

We study a non-relativistic system characterized by the Bose-Hubbard Lagrangian 
\begin{eqnarray}
  {\cal L}(n,t)
  &=& \frac{i}{2}\left[\Psi_n^*(t)\partial_{t}\Psi_n(t)
                     -\Psi_n(t)\partial_{t}\Psi_n^*(t)\right]
  \nonumber\\
  &&\
     +J[\Psi_n^*(t)\Psi_{n+1}(t)+\Psi_{n+1}^*(t)\Psi_{n}(t)]
  \nonumber\\
  &&\
     -\epsilon_n\Psi_n^*(t)\Psi_n(t)
     -\frac{U}{2}[\Psi_n^*(t)\Psi_n(t)]^2.
\label{eq:ClassLD}
\end{eqnarray}
Here, $\Psi$ is a complex-valued, i.e. two-component, field defined in time $t$ and at site $n$ of the $1$-dimensional lattice.
We use units where $\hbar=1$.
$J$ is the coupling which depends on the hopping probability and therefore on the lattice depth, $\epsilon_i$ denotes an external potential, and $U$ a real-valued coupling constant.
Only local interactions, between atoms at one lattice site are taken into account.

\subsection{Generating functional for correlation functions}
\label{sec:GenFunc}
The functional integral formulation of the quantum dynamics of a Bose-Einstein gas has been discussed extensively in the literature.
For details we refer to our recent discussion \cite{Gasenzer2005a} of the non-perturbative effective action approach to far-from-equilibrium dynamics of an ultracold Bose gas.
To simplify our notation we will, in the following, change to a representation of the field $\Psi$ in terms of its real and imaginary parts, i.e.,
\begin{eqnarray}
  &&\Phi_1 = \sqrt{2}\, \mathrm{Re}\Psi,\qquad 
  \Phi_2 = \sqrt{2}\, \mathrm{Im}\Psi,
  \label{eq:12basis}
\end{eqnarray}
such that $\Psi=(\Phi_1+i\Phi_2)/\sqrt{2}$.
Furthermore, we include the lattice index $n$ into the argument of the field $\Phi_i(x)$, i.e., $x=(t,n)\equiv(x_0,n)$.

The time evolution of quantum correlation functions can be derived from a generating functional.
This involves fields $\hat\Phi_i$ obeying Bose commutation relations $[\hat\Phi_i(t,\xv),\hat\Phi_j(t,\yv)]=-\sigma_{2,ij}\delta(\xv-\yv)$, where $\sigma_2$ denotes the Pauli 2-matrix, as well as the non-equilibrium nor\-malized initial-state density matrix $\hat{\rho}_D(t_0)$:  
\begin{eqnarray}
\label{eq:NEgenFuncZ}
  Z[J,K;\hat\rho_D]
  &=& \mathrm{Tr}\Big[\hat\rho_D(t_0)\,{\cal T}_{\cal C}
     \exp\Big\{i
     \int_{x,\cal C}\,\hat\Phi_i(x)J_i(x)
     \nonumber\\
  &&\
     +\frac{i}{2}\int_{xy,\cal C}\,\hat\Phi_i(x)K_{ij}(x,y)\hat\Phi_j(y)\Big\}\Big],
\end{eqnarray} 
where $\int_x\equiv\int_0^\infty dx_0\sum_n$.
Summation over double indices, e.g. $i=1,2$, is implied.
${\cal T}_{\cal C}$ denotes time ordering of the exponential integral along the closed time path $\cal C$ which extends from the initial time $x_0=0$ to the largest relevant time in the functional integral and back to $x_0=0$, i.e., our short hand notation means
\begin{eqnarray}
  \int_{x,\cal C} 
  = \int_{\cal C}dx_0\,\sum_n
  = \left(\int_0^Tdx_0+\int_T^0dx_0\right)\sum_n.
\label{eq:CTPInt}
\end{eqnarray}
Note that the closed-time contour can be seen as ensuring the normalization of the generating functional $Z$.
The terms linear in the external fields $J_i$ have been added to allow to generate correlation functions of order $n$ by functional differentiation:
\begin{eqnarray}
  \langle {\cal T_C}\hat\Phi(x_1)\cdots\hat\Phi(x_n)\rangle
  &=& \left.\frac{\delta^nZ[J,K;\hat\rho_D]}{i^n\delta J(x_1)\cdots\delta J(x_n)}
     \right|_{J=K\equiv0},
\label{eq:ClCorrFfromZ}
\end{eqnarray}
where the field indices have been suppressed.
The fields $K_{ij}$ are needed for the self-consistency condition of the two-point correlation functions as described below.

The generating functional (\ref{eq:NEgenFuncZ}) can be written, for Gaussian initial states, for which all correlation functions of order $n\ge3$ vanish at the initial time, as the functional integral (cf., e.g., \cite{Berges2004a})
\begin{align}
\label{eq:ZQuant}
  Z[J,K] 
  &= \int{\cal D}\Phi\,\exp\Big\{i\Big[
     S[\Phi] + \int_{x,\cal C}\,\Phi_i(x)J_i(x)
     \nonumber\\
  &\qquad
     +\frac{1}{2}\int_{xy,\cal C}\,\Phi_i(x)K_{ij}(x,y)\Phi_j(y)\Big]\Big\}.
\end{align} 
Here, 
\begin{eqnarray}
  S[\Phi]
  &=& \int_{xy,\cal C} \Bigg[\frac{1}{2}\Phi_i(x)\,iD_{ij}^{-1}(x,y)\Phi_j(y)
  \nonumber\\
  &&\
  -\frac{U}{4\cal N}\,\delta_{\cal C}(x,y)\,\Phi_i(x)\Phi_i(x)\Phi_j(x)\Phi_j(x)\Bigg],
\label{eq:Sq}
\end{eqnarray}
is the classical action depending on the fluctuating fields $\Phi_i$.
The number of field components is ${\cal N}=2$, here.
The delta function is defined as $\delta_{\cal C}(x,y)=\delta_{nm}\delta_{\cal C}(x_0-y_0)$, with $x=(x_0,n)$, $y=(y_0,m)$.
The classical inverse propagator reads
\begin{eqnarray}
  iD_{ij}^{-1}(x,y)
  &=&-i\delta_{\cal C}(x,y)\sigma_{2,ij}\partial_{y_0}
  \nonumber\\
  &&-\ H_\mathrm{1B}(x,y)
     \delta_{\cal C}(x_0-y_0)\delta_{ij},
\end{eqnarray}
where
\begin{eqnarray}
  &&H_\mathrm{1B}(x,y)=-J(\delta_{n+1,m}+\delta_{n,m+1})+\epsilon_n\delta_{nm}
\end{eqnarray}
is the one-body Hamiltonian.
Note that the Gaussian initial conditions have been absorbed into the current terms by redefining the fields $J$ and $K$ accordingly \cite{Berges2004a}.

\subsection{Effective action and dynamic equations}
\label{sec:EAandDynEq}
In most situations the correlation functions up to order $n=2$ are of primary interest, i.e., the mean field $\phi_i(x)=\langle\Phi_i(x)\rangle$ and the normal and anomalous one-body density matrices, which, in the basis (\ref{eq:12basis}) relate to the diagonal ($i=j$) and off-diagonal ($i=3-j$) matrix elements of the two-point functions $G_{ij}(x,y)=\langle{\cal T}_{\cal C}\hat{\Phi}_i(x)\hat{\Phi}_j(y)\rangle_c$, here generalized to unequal times $x_0\not=y_0$\footnote{%
The subscript $c$ indicates that $G$ is a connected Green's function or cumulant, $\langle{\cal T}_{\cal C}\Phi_i(x)\Phi_j(y)\rangle_c=\langle{\cal T}_{\cal C}\Phi_i(x)\Phi_j(y)\rangle-\phi_i(x)\phi_j(y)$.}.
In Ref.~\cite{Gasenzer2005a} we have recalled the essential aspects of the 2PI (2-particle-irreducible) effective action approach which allows in a transparent way to obtain equations of motion for these one- and two-point Green's functions within perturbative as well as non-perturbative approximation schemes with respect to the interaction constant $U$.
We summarize the main results in the following.

Starting from the generating functional $Z[J,K]=\exp\{iW[J,K]\}$, Eq.~(\ref{eq:ZQuant}), one derives, by double Legendre transform of the generating functional $W[J,K]$ of connected Green's functions, the 2PI effective action which is a functional of the one- and two-point functions \cite{Luttinger1960a,Baym1962a,Cornwall1974a}:
\begin{align}
\label{eq:2PIEAexp}
  \Gamma[\phi,G]
  &= S[\phi] +\frac{i}{2}\,\Tr\left\{\ln G^{-1}+G_0^{-1}[\phi]G\right\} 
     +\Gamma_2[\phi,G]\nonumber\\
  &\qquad+\mathrm{const.}\,
\end{align} 
Here,
\begin{eqnarray}
\label{eq:G0invexpl}
  &&iG_{0,ij}^{-1}(x,y;\phi)
  = \frac{\delta^2S[\phi]}{\delta\phi_i(x)\delta\phi_j(y)}
  = \delta_{\cal C}(x_0-y_0)
  \nonumber\\
  &&\quad\times\ \Bigg(iD^{-1}_{ij}(x,y)
     -\frac{2U}{\cal N}\big[\phi_i(x)\phi_j(x)
  \nonumber\\
  &&\qquad\quad
    +\ \phi_k(x)\phi_k(x)
      \delta_{ij}/2\big]\delta_{nm}\Bigg)
\end{eqnarray} 
is the classical inverse propagator, with $x=(x_0,n)$, $y=(y_0,m)$.
$\Gamma[\phi,G]$ describes the quantum system completely, i.e., knowing it, one derives, by means of Hamilton's principle of stationary action,
\begin{eqnarray}
\label{eq:2PIStatCondPhi}
  &&\frac{\delta\Gamma[\phi,G]}{\delta \phi_i(x)} = 0,\qquad \mbox{and}\qquad
  \frac{\delta\Gamma[\phi,G]}{\delta G_{ij}(x,y)} = 0,
\end{eqnarray} 
the exact many-body time evolution equations for $\phi$ and $G$.

As originally discussed in Refs.~\cite{Luttinger1960a,Baym1962a,Cornwall1974a}  $\Gamma_2[\phi,G]$ is represented as the sum of all closed two-particle irreducible (2PI) diagrams involving only the field $\phi$, the full propagator $G$ and bare vertices $\propto U/\cal N$ \footnote{%
In the context of condensed matter physics, a theory derivable from such an action functional is termed $\Phi$-derivable.}.
To solve the dynamic equations approximations are made on the level of the effective action, by truncating the diagrammatic expansion of $\Gamma_2[\phi,G]$.
An important advantage of this approach is that crucial symmetries like total particle number, energy, momentum, angular momentum, etc., are automatically fulfilled irrespective of the particular truncation, cf., e.g.~Ref.~\cite{Gasenzer2005a}.
In the appendix, we provide an explicit expression for the total energy within the NLO $1/\cal N$ approximation employed in this work.

The equations of motion resulting from Eqs.~(\ref{eq:2PIStatCondPhi}) are most conveniently written in terms of the real valued statistical and spectral correlation functions, 
\begin{eqnarray}
\label{eq:Fij}
  F_{ij}(x,y)
  &=&\frac{1}{2}\langle \{\hat{\Phi}_i(x),\hat{\Phi}_j(y)\}\rangle_c,
  \\
\label{eq:rhoij}
  \rho_{ij}(x,y)
  &=&i\langle [\hat{\Phi}_i(x),\hat{\Phi}_j(y)] \rangle_c,
\end{eqnarray} 
respectively.
These are related to $G$ through 
\begin{align}
  &G_{ij}(x,y)
  = F_{ij}(x,y) -\frac{i}{2}\rho_{ij}(x,y)\,\mathrm{sgn}_{\cal C}(x_0-y_0),
\end{align} 
where $\mathrm{sgn}_{\cal C}$ is the sign function which evaluates to $1$ for $x_0\ge y_0$ and to $-1$ otherwise.
In this representation, the time ordering translates into time integration limits in the equations of motion:
%
\begin{eqnarray}
\label{eq:EOMphi}
   &&\Big(-i\sigma_{2,ij}\partial_{x_0}
    -\frac{2U}{\cal N}\,F_{ij}(x,x)\Big)\phi_j(x)
   \nonumber\\
   &&\
    -\int_{y,\cal C}\Big(H_\mathrm{1B}(x,y)
    +\frac{U}{\cal N}\big[\phi_k(x)\phi_k(x)
   \nonumber\\
   &&\qquad\qquad\qquad\qquad\qquad
   +\ F_{kk}(x,x)\big]
    \delta_{\cal C}(x,y)\Big)
    \phi_i(x)
   \nonumber\\
   &&\quad
   = \int_0^{x_0}dy\,\Sigma^\rho_{ij}(x,y;\phi\equiv 0)\,\phi_j(y),
   \\
\label{eq:EOMFrho}
   &&\int_{z,\cal C}\big[-i\sigma_{2,ik}\delta_{\cal C}(x,z)\partial_{z_0}
    -M_{ik}(x,z)\big]\,
    \left(\begin{array}{r}F_{kj}(z,y) \\ \rho_{kj}(z,y) \end{array}\right)
   \nonumber\\
   &&\quad
    = \int_0^{x_0}dz\,\Sigma^\rho_{ik}(x,z;\phi)
      \left(\begin{array}{r}F_{kj}(z,y) \\ \rho_{kj}(z,y) \end{array}\right)
    \nonumber\\
   &&\quad\qquad
     -\int_0^{y_0}dz\,
      \left(\begin{array}{r}\Sigma^F_{ik}(x,z;\phi) \\ 
                            \Sigma^\rho_{ik}(x,z;\phi) 
			    \end{array}\right)\rho_{kj}(z,y).
\end{eqnarray} 
%
Here, 
\begin{align}
\label{eq:MofF}
  &M_{ij}(x,y)
   = \delta_{ij}
  \Big[H_\mathrm{1B}(x,y)
  \nonumber\\
  &\qquad
   +\ \frac{U}{\cal N}\Big(\phi_k(x)\phi_k(x)+F_{kk}(x,x)\Big)
     \delta_{\cal C}(x,y)\Big]
  \nonumber\\ 
  &\qquad
   + \frac{2U}{\cal N}\Big(\phi_i(x)\phi_j(x)+F_{ij}(x,x)\Big)\delta_{\cal C}(x,y).
\end{align} 
is the mean-field energy matrix which includes the $\phi_i$-dependent terms of the classical inverse propagator $iG_{0,ij}^{-1}$, Eq.~(\ref{eq:G0invexpl}), and the local part $\Sigma_{ij}^{(0)}(x)$ of the self energy
\begin{align}
\label{eq:Sigma}
  \Sigma_{ij}(x,y;\phi)
  &= {2i}\frac{\delta\Gamma_2[\phi,G]}{\delta G_{ij}(x,y)}
\end{align} 
which, to derive the above equations, has been decomposed into real and imaginary parts as $\Sigma_{ij}(x,y;\phi)=\Sigma_{ij}^{(0)}(x)\delta_{\cal C}(x,y)+\Sigma^F_{ij}(x,y;\phi)-(i/2)\Sigma^\rho_{ij}(x,y;\phi)\,\mathrm{sgn}_{\cal C}(x_0-y_0)$ \cite{Gasenzer2005a}.

We point out that, neglecting the right-hand sides of Eqs.~(\ref{eq:EOMphi}), (\ref{eq:EOMFrho}), these equations constitute a set of time-dependent Hartree-Fock-Bogoliubov (HFB) equations for the mean field and the two-point functions, cf., e.g. Refs.~\cite{Rey2004a,Gasenzer2005a}.
In this approximation, the dynamics of $\rho$ decouples from that of $\phi$ and $F$. 
Neglecting also $F$, Eq.~(\ref{eq:EOMphi}) is the Gross-Pitaevskii equation.
Given the exact self energy $\Sigma$, Eqs.~(\ref{eq:EOMphi}) and (\ref{eq:EOMFrho}) are the exact equations for the field $\phi$ and the correlation functions $F$ and $\rho$, respectively.
Eq.~(\ref{eq:EOMFrho}) is equivalent to the Schwinger-Dyson equation for the full Green's function $G$.

\subsection{Non-perturbative $1/{\cal N}$ approximation}
\label{sec:1overNapprox}
\begin{figure}[tb]
\begin{center}
\includegraphics[width=0.45\textwidth]{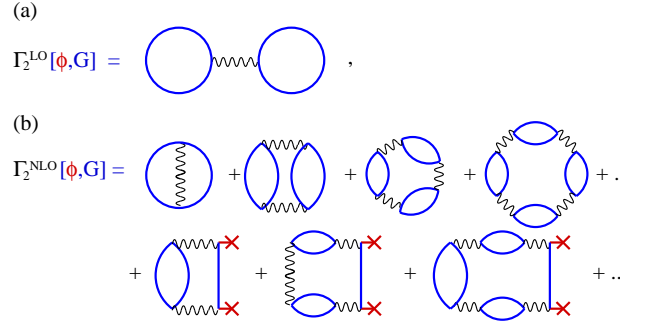}
\end{center}
\vspace*{-3ex}
\caption{
(Color online) Diagrammatic representation of the  leading order (LO) and next-to-leading order (NLO) contributions in the $1/\cal N$-expansion, to the 2PI part $\Gamma_2[\phi,G]$ of the 2PI effective action.
The thick blue lines represent 2-point functions $G_{ij}(x,y)$, the red crosses field insertions $\phi_i(x)$, and the wiggly lines vertices $U\delta_{\cal C}(x,y)$.
At each vertex, it is summed over double field indices $i$ and integrated/summed over double time and space variables $x$.
}
\label{fig:DiagrExpGamma2NLO}
\end{figure}

To derive the quantum many-body time evolution, details about the self energy $\Sigma$ are required, and these are, in general, only available to a certain approximation. 
In the following we will consider an expansion of $\Gamma_2$, to next-to-leading order (NLO) in powers of the inverse number of field components $\cal N$ \cite{Berges2002a,Aarts2002b,Berges2005a} which, in our special case, is ${\cal N}=2$.
In the context of a non-relativistic Bose gas, this approximation has been discussed in detail in Ref.~\cite{Gasenzer2005a}.
The contribution $\Gamma_2[\phi,G]$ to the effective action $\Gamma$, Eq. (\ref{eq:2PIEAexp}), then involves a leading (LO) and next-to-leading order (NLO) part which can be diagrammatically represented as shown in Fig.~\ref{fig:DiagrExpGamma2NLO}.
While the leading order contribution involves one diagram, in NLO a chain of bubble diagrams is resummed.
All of these diagrams are proportional to the same power of $1/\cal N$ since each vertex scales with $1/\cal N$, which is cancelled by the (blue) propagator loops which scale with $\cal N$ since they involve a summation over the field indices from $1$ to $\cal N$.
Note that the Hartree-Fock-Bogoliubov (HFB) approximation is given by an action $\Gamma_2$ which involves $\Gamma_2^\mathrm{LO}$ and the first diagram of $\Gamma_2^\mathrm{NLO}$ in Fig.~\ref{fig:DiagrExpGamma2NLO} (b), cf., e.g., \cite{Gasenzer2005a}.
We compare, in the next section, our results in NLO of the $1/\cal N$ approximation with the HFB dynamics as well as with a perturbatively%
\footnote{The term `perturbative' is used at the level of the 2PI effective action and therefore refers to a finite number of bare vertices in a certain 2PI diagram, irrespective of the inherently non-perturbative character of the self-consistently determined two-point functions.}
reduced $1/\cal N$ approximation, the second-order coupling expansion, which, besides the HFB diagrams, takes into account still the two diagrams of second-order in $U$, i.e. the 2nd and 5th diagrams in Fig.~\ref{fig:DiagrExpGamma2NLO} (b) which contain two wiggly lines each.

From $\Gamma_2[\phi,G]=\Gamma_2^\mathrm{LO}[\phi,G]+\Gamma_2^\mathrm{NLO}[\phi,G]$ we obtain, using Eq.~(\ref{eq:Sigma}), the self energies $\Sigma_{ij}(x,y)=\Sigma^F_{ij}(x,y)-(i/2)\mathrm{sgn}_{\cal C}(x_0-y_0)\Sigma^\rho_{ij}(x,y)$, with
\begin{align}
  &\left(\begin{array}{r}
        \Sigma^F_{ij}(x,y) \\ -\frac{1}{2}\Sigma^\rho_{ij}(x,y)
	\end{array}\right)
   = -\frac{2U}{\cal N}\Bigg[\left(\begin{array}{r}
        I_F(x,y) \\ -\frac{1}{2}I_\rho(x,y)
	\end{array}\right)
        \phi_i(x)\phi_j(y) 
  \nonumber\\
  &\quad 
  +\ \left(\begin{array}{rr}
        \Delta_F(x,y) & 
	\frac{1}{2}\Delta_\rho(x,y) \\
       -\frac{1}{2}\Delta_\rho(x,y) & 
        \Delta_F(x,y) \end{array}\right)
    \left(\begin{array}{r}
        F_{ij}(x,y) \\ -\frac{1}{2}\rho_{ij}(x,y)
	\end{array}\right)\Bigg],
\label{eq:SigmaNLO1N}
\end{align}
where $\Delta_{F,\rho}(x,y)=I_{F,\rho}(x,y)+P_{F,\rho}(x,y;I_{F,\rho})$.
The resummation to NLO in $1/\cal N$ is expressed by the coupled integral equations for $I_{F,\rho}$ \cite{Berges2005a}%
\footnote{Concerning Eqs.~(B5-B8) and (B13-B14) of Ref.~\cite{Gasenzer2005a} an erratum is to be published.}:
\begin{align}
  &\left(\begin{array}{r}
        I_F(x,y) \\ I_\rho(x,y)
	\end{array}\right)
  =
  \frac{U}{\cal N}\Bigg[
  \left(\begin{array}{c}
        F(x,y)^2-\frac{1}{4}\rho(x,y)^2 \\
	2F_{ij}(x,y)\rho_{ij}(x,y)
	\end{array}\right)
  \nonumber\\
   &\quad
   -\ \int_0^{x_0}dz\,I_\rho(x,z)
      \left(\begin{array}{c}
            F(z,y)^2-\frac{1}{4}\rho(z,y)^2 \\ 
            2F_{ij}(z,y)\rho_{ij}(z,y) 
	    \end{array}\right)
    \nonumber\\
   &\quad
   +\ \int_0^{y_0}dz\,
      \left(\begin{array}{r}
            I_F(x,z) \\ 
            I_\rho(x,z) 
	    \end{array}\right)2F_{ij}(z,y)\rho_{ij}(z,y)\Bigg].
\label{eq:IFrho}
\end{align}
Here, $F^2=F_{ij}F_{ij}$, etc.
The functions $P_{F,\rho}$, which contribute to $\Delta_{F,\rho}$ in the self energies (\ref{eq:SigmaNLO1N}) and vanish if $\phi_i\equiv0$, read \cite{Berges2005a}
\begin{align}
  &P_F(x,y;I_{F,\rho})
  = -\frac{2U}{{\cal N}}\Big\{
     H_F(x,y)
  \nonumber\\
  &\ 
  +\int_0^{y_0}dz\left[H_F(x,z)I_\rho(z,y)+I_F(x,z)H_\rho(z,y)\right]
  \nonumber\\
  &\ 
  -\int_0^{x_0}dz\left[H_\rho(x,z)I_F(z,y)+I_\rho(x,z)H_F(z,y)\right]
  \nonumber\\
  &\ 
  -\int_0^{x_0}dv\int_0^{y_0}dw\, I_\rho(x,v)H_F(v,w)I_\rho(w,y)
  \nonumber\\
  &\ 
  +\int_0^{x_0}dv\int_0^{v_0}dw\, I_\rho(x,v)H_\rho(v,w)I_F(w,y)
  \nonumber\\
  &\ 
  +\int_0^{y_0}dv\int_{v_0}^{y_0}dw\, I_F(x,v)H_\rho(v,w)I_\rho(w,y)\Big\},
\label{eq:PF}
\end{align}
\begin{align}
  &P_\rho(x,y;I_{F,\rho})
  = -\frac{2U}{{\cal N}}\Big\{
     H_\rho(x,y)
  \nonumber\\
  &\ 
  -\int_{y_0}^{x_0}dz\left[H_\rho(x,z)I_\rho(z,y)+I_\rho(x,z)H_\rho(z,y)\right]
  \nonumber\\
  &\ 
  +\int_{y_0}^{x_0}dv\int_{y_0}^{v_0}dw\, I_\rho(x,v)H_\rho(v,w)I_\rho(w,y)\Big\},
\label{eq:Prho}
\end{align}
wherein the functions $H_{F,\rho}$ are defined as
\begin{align}
  H_F(x,y)
  &= -\phi_i(x)F_{ij}(x,y)\phi_j(y),
  \nonumber\\
  H_\rho(x,y)
  &= -\phi_i(x)\rho_{ij}(x,y)\phi_j(y).
\label{eq:HFHrho}
\end{align}
The technical procedure to solve the above dynamic equations in every time step requires the determination of the 
functions $I(x,y)$, before the actual propagation of the respective correlation functions.

\section{Far-from-equilibrium time evolution of a one-dimensional lattice Bose gas}
\label{sec:NumRes}
\subsection{Initial conditions}
In the following we present our results for the time evolution of a lattice Bose gas initially in a state far from equilibrium.
The 2PI approach invoked in this work allows us to specify initial states which are Gaussian in the sense that only the mean-field and 2-point correlation functions are non-zero at the initial time $t_0=0$.
Note that, using a general nPI approach \cite{Berges2004a}, also initial states with higher-order correlations can be taken into account.

The total number of particles $N_{\mathrm{tot},n}(t)$ at lattice site $n$ and time $t$ is given as the sum of the number of condensed and excited atoms ($x=(t,n)$):
\begin{align}
  N_{\mathrm{tot},n}(t)
  &= \langle\hat{\Psi}^\dagger(x)\hat{\Psi}(x)\rangle
  =N_{\mathrm{c},n}(t)+N_{\mathrm{exc},n}(t),
\label{eq:ntot}
  \\
  N_{\mathrm{c},n}(t) 
  &=\frac{1}{2}\left[\phi_i(x)\phi_i(x)\right],
\label{eq:nc}
  \\
  N_{\mathrm{exc},n}(t) 
  &=\frac{1}{2}\left[F_{ii}(x,x)-1\right],
\label{eq:nexc}
\end{align}
where the constant $-1$ inside the parantheses takes into account the zero-point fluctuations.
Formally, this $-1$ originates from the spectral function $\rho$ which, due to the Bose commutation relations, vanishes at equal times, except for
\begin{align}
  &-\rho_{12}(x_0,n;x_0,m)
    =\rho_{21}(x_0,n;x_0,m)
    =\delta_{nm}.
\label{eq:rhoatequaltimes}
\end{align}
By virtue of the 2PI effective action approach, total particle number as well as total energy are exactly conserved in all approximations obtained by truncating the diagrammatic expansion of $\Gamma_2$, cf.~\ref{app:totenergy}.

We consider initial states where the normal and anomalous fluctuations are zero, i.e., all atoms are in the condensate: $\langle\hat{\Psi}^\dagger(x)\hat{\Psi}(y)\rangle_c=[F_{ii}(x,y)-\delta_{\cal C}(x,y)]/2=0$, and $\langle\hat{\Psi}(x)\hat{\Psi}(y)\rangle_c=[F_{11}(x,y)-F_{22}(x,y)]/2+iF_{12}(x,y)=0$, for $x_0=y_0=0$.
Hence we choose
\begin{eqnarray}
  F_{11}(0,n;0,m) = F_{22}(0,n;0,m)
  &=& \delta_{nm}/2,
  \nonumber\\
  F_{12}(0,n;0,m) = F_{21}(0,n;0,m)
  &=& 0.
\label{eq:Fini}
\end{eqnarray}
The initial values for $\rho_{ij}$ are prescribed by Eq.~(\ref{eq:rhoatequaltimes}).
The condensate fraction is chosen to be non-zero at a single site $m$ only,
\begin{eqnarray}
  \phi_j(0,n)
  &=& \sqrt{2N}\,\delta_{j1}\delta_{nm},
\label{eq:}
\end{eqnarray}
where $N$ is the total number of particles, and $j=1$ implies that the initial mean field $\psi=(\phi_1+i\phi_2)/\sqrt{2}$ is real.

\subsection{Numerical results}
%
\begin{figure}[tb]
\begin{center}
\includegraphics[width=0.48\textwidth]{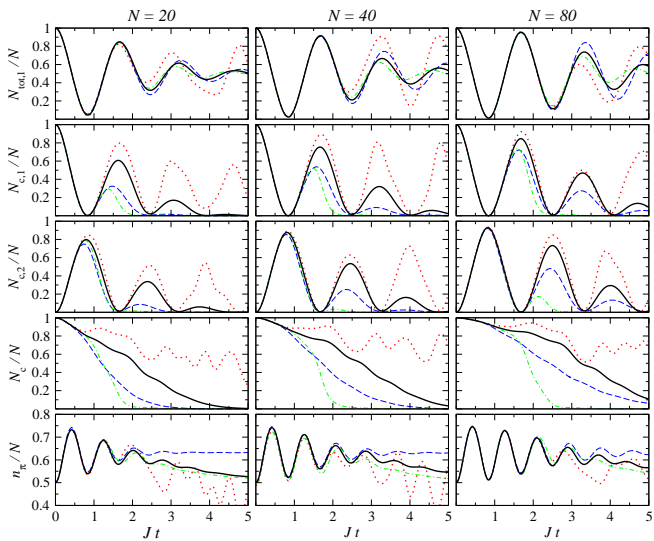}
\end{center}
\vspace*{-3ex}
\caption{
(Color online) Time evolution of an ultracold Bose gas in a 1-dimensional lattice with $N_\mathrm{L}=2$ lattice sites and periodic boundary conditions.
The three columns show the same quantities, for a different total number of atoms $N=20$, $40$, and $80$, respectively.
$U/J$ is chosen such that the characteristic interaction parameter is $NU/J\equiv4$ throughout.
Initially, all atoms are in a Bose-Einstein condensate at lattice site $1$.
The graphs in the first line show the total number of atoms $N_\mathrm{tot,1}$ at site $1$, normalized to $N$.
Due to number conservation, one has $N_\mathrm{tot,2}=N-N_\mathrm{tot,1}$.
The second and third lines show the condensate fractions $N_{\mathrm{c},n}/N$ at sites $n=1,2$, and the fourth line adds these up, $N_\mathrm{c}/N=\sum_nN_{\mathrm{c},n}/N$.
The last line shows the occupation $n_q$ of the quasimomentum $q=\nu\pi$ [in units of $1/$Lattice spacing] (\ref{eq:qmom}), for $\nu=1$, normalized to the total quasimomentum occupation $n_0+n_\pi=N$.
Notice the different scale in the last line.
Our results from the $1/\cal N$ expansion to NLO are shown as a thick solid line.
The (blue) dashed lines show the results from an exact calculation by A.M.~Rey and coworkers, cf.~Figs.~5, 6 of Ref.~\cite{Rey2004a}.
The (green) dash-dotted curves show the dynamics resulting from a $1/\cal N$ expansion in perturbative 2nd-order coupling approximation (see text), the (red) dotted curves correspond to the Hartree-Fock-Bogoliubov approximation.
With both sets of curves we reproduce the respective results in \cite{Rey2004a}. 
}
\label{fig:dynamics2}
\end{figure}
\begin{figure}[tb]
\begin{center}
\includegraphics[width=0.48\textwidth]{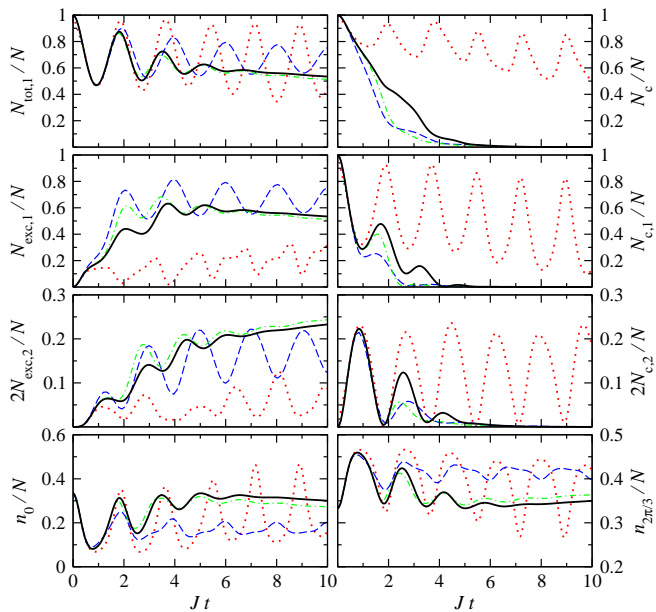}
\end{center}
\vspace*{-3ex}
\caption{
(Color online) Time evolution of an ultracold Bose gas in a 1-dimensional lattice with $N_\mathrm{L}=3$ lattice sites and periodic boundary conditions.
Only site $n=1$ is filled initially with $N=8$ atoms.
The interaction strength is $U/J=1/3$, such that $NU/J=2.67$.
Besides the quantities already shown in Fig.~\ref{fig:dynamics2}, we show the depletion $N_{\mathrm{exc},n}/N=(N_{\mathrm{tot},n}-N_{\mathrm{c},n})/N$ at sites $n=1,2$. 
The populations of site $n=3$ are the same as those at $n=2$ due to the periodic boundary conditions.
The curves indicate the same approximations as in Fig.~\ref{fig:dynamics2}.
Notice the different scale in the third and fourth lines. 
}
\label{fig:dynamics3}
\end{figure}

The expansion in inverse powers of the number of field components $\cal N$ allows for approximations which are inherently non-perturbative, i.e., it takes into account diagrams up to infinite order in the physical coupling constant $U$.
Dividing the equations of motion by the tunneling parameter $J$, neglecting quantum fluctuations%
\footnote{It can be shown that quantum statistical fluctuations are taken into account by the $\rho^2$ terms in the functions $I_{F,\rho}$ as well as the $\Delta_\rho\rho_{ij}$ term in $\Sigma_{ij}^F$ \cite{Gasenzer2006a}.}, 
and taking into account that the $F_{ij}(x,y)$ and $\phi_i(x)\phi_j(y)$ approximately scale with the total number of particles $N$, one finds that all terms in the NLO $1/\cal N$ equations of motion (\ref{eq:EOMphi}), (\ref{eq:EOMFrho}) scale with some power of $NU/J$.
In the series representing the self energy $\Sigma_{ij}$, arbitrary high powers are resummed.
The remaining terms which take into account the quantum fluctuations \cite{Gasenzer2006a} are reduced by powers of $N$ since the spectral functions $\rho_{ij}(x,y)$ are of order one.
Nevertheless, if $NU/J>1$ one expects that any perturbative or loop approximation of $\Gamma_2$, neglecting terms with higher powers of $NU/J$, fails.
In order to probe the accuracy of the non-perturbative $1/\cal N$ approximation we chose $NU/J$ to be larger than $1$ and compare our results with the results of an exact numerical calculation \cite{Rey2004a} for limited numbers of lattice sites and particles.
Following Ref.~\cite{Rey2004a} we have considered cases with a total number of $N_\mathrm{L}=2$ and $N_\mathrm{L}=3$ lattice sites, and particle numbers between $N=8$ and $80$.
Periodic boundary conditions were imposed.
In Figs.~\ref{fig:dynamics2} and \ref{fig:dynamics3} we present our results for the non-equilibrium quantum many-body evolution in the non-perturbative next-to-leading order (NLO) $1/\cal N$ approximation as well as the `perturbative' Hartree-Fock-Bogoliubov (HFB) and second-order coupling approximations.
In Fig.~\ref{fig:dynamics2}, we consider 2 lattice sites.
The graphs in the first line show the total number of atoms $N_\mathrm{tot,1}$ at site $1$, normalized to $N$.
Due to number conservation, one has $N_\mathrm{tot,2}=N-N_\mathrm{tot,1}$.
The second and third lines show the condensate fractions $N_{\mathrm{c},n}/N$ at sites $n=1,2$, cf.~Eq.~(\ref{eq:nc}), and the fourth line adds these up, $N_\mathrm{c}/N=\sum_nN_{\mathrm{c},n}/N$.
The last line shows the occupation $n_{q_\nu}$, of the quasimomentum $q_\nu=2\pi\nu/N_\mathrm{L}$, $\nu=1$ [$q$ in units of $1/$Lattice spacing], normalized to the total quasimomentum occupation $\sum_{\nu=0}^{N_\mathrm{L}}n_{q_\nu}=N$.
$n_q$ is defined as
\begin{align}
  &n_q(t)
  =\frac{1}{N_\mathrm{L}}\sum_{nm}e^{iq(n-m)}\,
     \langle\Psi^\dagger_n(t)\Psi_m(t)\rangle
  \nonumber\\
  &=\frac{1}{2N_\mathrm{L}}\sum_{nm}\Big\{
     \cos[q(n-m)]\Big[F_{kk}(x,y)+\phi_k(x)\phi_k(y)\Big]
  \nonumber\\
  &\quad\qquad\qquad
     +\sin[q(n-m)]\Big[F_{21}(x,y)-F_{12}(x,y)
  \nonumber\\
  &\qquad\qquad\qquad
                      +\phi_2(x)\phi_1(y)-\phi_1(x)\phi_2(y)\Big]-1\Big\},
\label{eq:qmom}
\end{align}
with $x=(t,n)$, $y=(t,m)$.
In Fig.~\ref{fig:dynamics3}, we consider 3 lattice sites, and plot the fractions of condensed, excited, and total atom numbers for sites 1 and 2, as well as the total number of condensed atoms $N_\mathrm{c}=N_{\mathrm{c},1}+2N_{\mathrm{c},2}$.
We also plot the quasimomentum populations $n_0$ and $n_{2\pi/3}$.
Note that the respective populations of sites 2 and 3 are equal due to the periodic boundary conditions.
For the quasimomentum populations this means that the term proportional $\sin[q(n-m)]$ in Eq.~(\ref{eq:qmom}) vanishes. 

The solid line shows, in all plots, the evolution according to the dynamic equations in NLO $1/\cal N$ approximation.
For comparison, we show, as a dashed line, the results of an exact calculation obtained by A.M.~Rey and coworkers for a coherent initial state of $N$ atoms at one site \cite{Rey2004a}.

The dotted and dash-dotted lines show the corresponding dynamics in the HFB and second-order coupling approximations, respectively, cf.~Sect.~\ref{sec:1overNapprox}.
In the HFB approximation all non-local self-energy terms vanish in the equations of motion takes (\ref{eq:EOMphi}) and (\ref{eq:EOMFrho}), and fluctuations enter in the form of a local energy shift only. 
This approximation is fully classical, i.e., it does not take into account any quantum statistical fluctuations, cf., e.g., Ref.~\cite{Gasenzer2006a} for a detailed discussion.
In the second-order coupling expansion of the NLO $1/\cal N$ approximation, non-local contributions from the self energy allow for damping.
In Figs.~\ref{fig:dynamics2} and \ref{fig:dynamics3}, the dotted and dash-dotted lines precisely confirm the respective results of Ref.~\cite{Rey2004a} for the HFB and 2$^\mathrm{nd}$-order $1/\cal N$ approximations. 

The time evolution shows different characteristic periods.
At early times, the condensate oscillates coherently between the lattice wells.
Only a small number of atoms is scattered from the condensate fraction into excited modes.
This dynamics is effectively described by a set of coupled non-linear Gross-Pitaevskii-like equations for the condensate mean field at the different lattice sites.
These have the form of coupled single-particle Schr\"odinger equations, and the dynamics corresponds to Rabi oscillations.
The frequency of these oscillations is determined by the expansion parameter $NU/J$.
In the case of 2 lattice wells, the Bose-Hubbard lattice gas with periodic boundary conditions resembles a Josephson junction with coupling energy $E_\mathrm{J}=2JN$,   and charging energy $E_\mathrm{c}=2U$, cf., e.g., \cite{Paraoanu2001a,Pitaevskii2001a}.
For vanishing $U$, the system undergoes Rabi oscillations with frequency $\Omega=2E_\mathrm{J}/N=4J$, i.e., the period of these oscillations is $T=(\pi/2)\,J^{-1}$.
Note also that the choice of parameters $NU/J=4$, $N^2>400$ sets the system into the Josephson regime $N^{-2}(NU/J)\ll(NU/J)\ll N^2$.
In this regime, the equilibrium state has a well defined relative phase between the sites, and small oscillations around this configuration can be described as a collective excitation, the Josephson plasmon with plasma frequency $\omega_\mathrm{JP}=\sqrt{E_\mathrm{J}(E_\mathrm{c}+4E_\mathrm{J}/N^2)}=2J\sqrt{(NU/J)+4}$ \cite{Paraoanu2001a}.
We find, however, that the frequency of the large oscillations in Fig.~\ref{fig:dynamics2} is closer to the Rabi frequency of an ideal gas than to the plasma frequency $\omega_\mathrm{JP}=4\sqrt{2}J$, for which $T_\mathrm{JP}\simeq0.35\pi/J$.
This is, of course, not contradictory since the derivation of the above expression for $\omega_\mathrm{JP}$ requires the assumption of small oscillations around equilibrium.

Due to the interactions between the atoms, also higher-order classical statistical and quantum correlations build up.
To leading order this means that atoms are exchanged between the condensate and the non-condensate modes of the gas.
These processes lead to a rapid destruction of the condensate fraction.
Hence, the coherence of the gas deteriorates which, on a somewhat longer timescale, leads to damping of the Rabi oscillations.

All three approximations describe the dynamics well within the first period of coherent oscillations.
This was expected since the system is initially mostly classical, with higher-order correlations and fluctuations playing a minor role.
At larger times, all approximations fail to describe the dynamics accurately.
While the purely classical HFB approxiation even qualitatively fails to show the correct damping behaviour, the higher-order approximations yield damping but quantitatively different results.
Our comparisons for different atom numbers $N$ indicate that the accuracy of the NLO $1/\cal N$ approximation improves with increasing $N$.
All examples show that the condensate fraction is, at large times, underestimated in the second-order coupling expansion.
Although it is overestimated in the NLO $1/\cal N$ approximation, our results show that the non-perturbative $1/\cal N$ resummation is capable of taking into account high correlations which, at large times become relevant.

\section{Conclusions and outlook}
\label{sec:Concl}
In this work we have studied the far-from-equilibrium dynamics of an ultracold, strongly interacting one-dimensional lattice Bose gas on the basis of the 2PI effective action in next-to-leading-order (NLO) $1/\cal N$ approximation.
The 2PI effective action preserves, at any truncation, vital conservation laws as total particle number and energy, and allows to derive in a consistent way non-perturbative approximations which remain applicable also for strong interactions.
For weakly interacting systems close to equilibrium, the 2PI approach gives the well-known mean-field theories and their kinetic extensions including the dynamics and dissipation of small excitations.
However, the 2PI approach goes substantially beyond the basic Hartree-Fock-Bogoliubov approximation and leads to a description of dynamical evolution far from equilibrium in the form of an initial-value problem.
Moreover, it is not restricted to a small number of particles.

The systems have been considered in the framework of the Bose-Hubbard model and consist of two and three lattice sites one of which is coherently populated initially while the others are empty.
Tunneling through the barriers leads to Rabi-like oscillations between the lattice wells which are damped due to particle interactions. 
We have studied the time evolution of the condensate and excited fractions of the gas as well as of the quasimomentum distribution and compared these results with those from the exact, fully quantal calculation of the dynamics from Ref.~\cite{Rey2004a}.
We find that in the non-perturbative $1/\cal N$ approximation the 2PI approach gives a reduced damping due to higher-order correlations as seen in the exact solution.
Quantitatively, the damping of the oscillations of the condensate fractions is underestimated in the NLO $1/\cal N$ approximation while it is overestimated for the excited fractions.
The agreement with the exact results appears to improve when choosing a larger total number $N$ of particles.
Our findings are similar to those in Ref.~\cite{Aarts2006a}, where the non-equilibrium dynamics of a single non-linear harmonic oscillator has been studied in the NLO as well as in a restricted NNLO $1/\cal N$ approximation and compared with the exact evolution.
As compared to these results, which were obtained in the symmetric phase where the mean field $\phi$ vanishes, we do not find, in NLO $1/\cal N$, irregular behaviour at large times, like the revivals seen in the Hartree-Fock-Bogoliubov approximation.
A systematic study of the $N$-dependence of the accuracy of the $1/\cal N$ approximation, extending the exact studies to a larger particle number and lattice size, using, e.g.,  stochastic quantization techniques along the lines of \cite{Berges2005b}, is the subject of ongoing work.
We furthermore expect that the particular choice of an initial coherent-state population affects the ensuing damping behaviour within the Markov time.
This requires a detailed study of the non-Markovian effects which shall be a further topic of a future publication.

\section{Acknowledgments}
\noindent
We are very grateful to J\"urgen Berges, Hrvoje Buljan, Markus Oberthaler, Michael G. Schmidt, J\"org Schmied\-mayer, Gora Shlyapnikov, and Amichay Vardi for valuable discussions, and to Werner Wetzel for his continuing support concerning computing facilities.
We would also like to thank Ana Maria Rey for allowing us to reproduce earlier results.
This work has been supported by the Deutsche Forschungsgemeinschaft (T.G.).

\begin{appendix}
 
\section{Number and energy conservation}
\label{app:totenergy}
\noindent
An important advantage of the 2PI effective action approach is that crucial symmetries like total particle number and energy are automatically fulfilled irrespective of the particular truncation.
As was shown in Ref.~\cite{Gasenzer2005a} number conservation is a consequence of the Noether theorem in conjunction with the invariance of the theory under transformations which are elements of the group $O({\cal N})$.
Each 2PI diagram in the expansion of $\Gamma[\phi,G]$ separately carries this property such that any truncation of the series leads to number conservation.
As a consequence, the total number 
\begin{eqnarray}
\label{eq:totaldensity}
  N(t)
  &=& \frac{1}{2}\int_{\vec{x}}\left[\phi_i(x)\phi_i(x)+G_{ii}(x,x)\right]
\end{eqnarray} 
is conserved in time \cite{Gasenzer2005a}.
For the lattice gas, the spatial integral means $\int_{\vec{x}}f(x)=\sum_n f(t,n)$.

Energy conservation follows from time translation invariance of $\Gamma$, cf., e.g., Ref.~\cite{Arrizabalaga2005a}.
Consider the general translations in continuous space and time which vanish at the boundary, $x^\mu\to x^\mu+\varepsilon^\mu(x)$, where $\varepsilon^\mu(x)$ is a time- and space-dependent infinitesimal 4-vector.
The mean field and 2-point functions transform, under these translations, to leading order in $\varepsilon$, as $\phi_i(x)\to\phi_i(x)+\varepsilon^\nu(x)\partial^x_\nu\phi_i(x)$, and $G_{ij}(x,y)\to G_{ij}(x,y)+\varepsilon^\nu(x)\partial^x_\nu G_{ij}(x,y)+\varepsilon^\nu(y)\partial^y_\nu G_{ij}(x,y)$, respectively.
Here, $\partial^x_\nu=\partial/\partial x^\nu$, etc.
One can show that under these transformations the variation of the 2PI effective action $\Gamma$ can be written as $\Gamma[\phi,G]\to\Gamma[\phi,G]+\delta\Gamma[\phi,G]$, with
\begin{eqnarray}
  \delta\Gamma[\phi,G] = \int_x T^{\mu\nu}(x)\,\partial^x_\mu\varepsilon_\nu(x).
\label{eq:deltaGamma}
\end{eqnarray}
Since, by virtue of the stationarity conditions (\ref{eq:2PIStatCondPhi}), the variation $\delta\Gamma$ vanishes for all solutions of the equations of motion for $\phi_i$ and $G_{ij}$, an integration by parts shows that $T^{\mu\nu}$ is the conserved Noether current for the time-space-translations:
\begin{eqnarray}
  \delta\Gamma[\phi,G] = -\int_x \varepsilon_\nu(x)\,\partial^x_\mu T^{\mu\nu}(x) = 0.
\label{eq:EMTconservation}
\end{eqnarray}
$T^{\mu\nu}(x)$ is identified as the energy-momentum tensor, and the conservation law for total energy is expressed as $\partial^x_\mu T^{\mu0}(x)=0$ or $\partial_t\int d^3x\,T^{00}(t,\vec{x})=0$.

To leading order in the $1/\cal N$ approximation we used the above space-time translations, and Eq.~(\ref{eq:deltaGamma}), to derive the energy-momentum tensor.
For the interaction terms which depend on the coupling $U$, however, as well as for the terms in NLO of the $1/\cal N$ approximation, it is more convenient to use a procedure known from field theory on curved space-time.
For a space-time-dependent metric $g_{\mu\nu}(x)$, the energy-momentum tensor is defined as \cite{Misner1973a}
\begin{eqnarray}
   T_{\mu\nu}(x)=\frac{2}{\sqrt{-g}}
   \frac{\delta\Gamma[\phi,G;g^{\mu\nu}]}{\delta g^{\mu\nu}},
\label{eq:EMTcurvedmetric}
\end{eqnarray}
where $\sqrt{-g}$ denotes the square root of minus the determinant of $g_{\mu\nu}$.

In the following we only quote the result for the total energy $E(t)=\sum_nT_{00}(t,n)$ of the lattice Bose gas described by the Lagrangian (\ref{eq:ClassLD}).
Using the above definitions and the useful relations $\delta g^{\mu\nu}=-g^{\mu\rho}g^{\nu\sigma}\delta g_{\rho\sigma}$ and $\delta\sqrt{-g}=\sqrt{-g}g^{\mu\nu}\delta g_{\mu\nu}/2$, one obtains, in flat Minkowski space-time, with $g^{\mu\nu}\equiv\mathrm{diag}\{1,-1,-1,-1\}$:
\begin{align}
  &E(t)
  =\frac{1}{2}\int_{\vec{x}y}\delta_{\cal C}(x,y)\Big\{
     H_\mathrm{1B}(x,y)\big[\phi_i(y)\phi_i(x)+G_{ii}(y,x)\big]
  \nonumber\\
  &\ +\frac{U}{\cal N}\left[\frac{1}{2}\left(\phi^2(x)+G_{ii}(x,x)\right)^2
                          -2H(x,x)\right]+ I(x,x) 
  \nonumber\\
  &\ +\frac{2U}{\cal N}\Big[
       2i\int_z H(x,z)I(x,z)
  \nonumber\\
  &\qquad\quad +\int_{zu}I(x,z)H(z,u)I(u,x)\Big]\Big\}. 
\label{eq:Etot}
\end{align}
Note that there is no integration over $x_0=t$.
Moreover, we use the definitions
\begin{eqnarray}
  I(x,y)
  &=&\frac{U}{\cal N}\left[G^2(x,y)-i\int_zI(x,z)G^2(z,y)\right],
\label{eq:I}
  \\
\label{eq:H}
  H(x,y)
  &=& -\phi_i(x)G_{ij}(x,y)\phi_j(y),
\end{eqnarray}
with $G^2=G_{ij}G_{ij}$.
The term proportional to $[G_{ii}(x,x)]^2$ in Eq.~(\ref{eq:Etot}) stems from the LO contribution in the expansion of $\Gamma_2$ in powers of $1/\cal N$, and the term $I(x,x)$ as well as the last line from the NLO contributions, cf.~Fig.~\ref{fig:DiagrExpGamma2NLO}.
\end{appendix}

\bibliography{bibtex/mybib,bibtex/additions}

\end{document}